\documentclass[aps,prl,twocolumn,superscriptaddress]{revtex4-2}
\usepackage{graphicx}
\usepackage{color}

\usepackage{mhchem}

\begin{document}
\title{Ultrafast carrier dynamics throughout the three-dimensional Brillouin zone of the Weyl semimetal PtBi$_2$}
\author{Paulina Majchrzak}
\affiliation{Department of Physics and Astronomy, Aarhus University, 8000 Aarhus C, Denmark}
\author{Charlotte Sanders}
\affiliation{Central Laser Facility, STFC Rutherford Appleton Laboratory, OX11 0QX, Harwell, UK}
\author{Yu Zhang}
\affiliation{Central Laser Facility, STFC Rutherford Appleton Laboratory, OX11 0QX, Harwell, UK}
\author{Andrii Kuibarov}
\affiliation{Leibniz IFW Dresden, 01069 Dresden, Germany}
\author{Oleksandr Suvorov}
\affiliation{Leibniz IFW Dresden, 01069 Dresden, Germany}
\author{Emma Springate}
\affiliation{Central Laser Facility, STFC Rutherford Appleton Laboratory, OX11 0QX, Harwell, UK}
\author{Iryna Kovalchuk}
\affiliation{Leibniz IFW Dresden, 01069 Dresden, Germany}
\affiliation{Kyiv Academic University, 03142 Kyiv, Ukraine}
\author{Saicharan Aswartham}
\affiliation{Leibniz IFW Dresden, 01069 Dresden, Germany}
\author{Grigory Shipunov}
\affiliation{Leibniz IFW Dresden, 01069 Dresden, Germany}
\author{Bernd B\"uchner}
\affiliation{Leibniz IFW Dresden, 01069 Dresden, Germany}
\affiliation{W\"urzburg-Dresden Cluster of Excellence ct.qmat, 01069 Dresden, Germany}
\author{Alexander Yaresko}
\affiliation{Max-Planck-Institute for Solid State Research, 70569 Stuttgart, Germany}
\author{Sergey Borisenko}
\affiliation{Leibniz IFW Dresden, 01069 Dresden, Germany}
\affiliation{W\"urzburg-Dresden Cluster of Excellence ct.qmat, 01069 Dresden, Germany}
\author{Philip Hofmann}
\email{philip@phys.au.dk}
\affiliation{Department of Physics and Astronomy, Aarhus University, 8000 Aarhus C, Denmark}

\begin{abstract}
Using time- and angle-resolved photoemission spectroscopy, we examine the unoccupied electronic structure and electron dynamics of the type-I Weyl semimetal PtBi$_2$. Using the ability to change the probe photon energy over a wide range, we identify the predicted Weyl points in the unoccupied three-dimensional band structure and we discuss the effect of $k_\perp$ broadening in the normally unoccupied states. We characterise the electron dynamics close to the Weyl point and in other parts of three-dimensional Brillouin zone using $k$-means, an unsupervised machine learning technique. This reveals distinct differences---in particular, dynamics that are faster in the parts of the Brillouin zone that host most of the bulk Fermi surface than in parts close to the Weyl points.
\end{abstract}
\maketitle

Dirac semimetals (DSMs) and Weyl semimetals (WSMs) are three-dimensional (3D) topological solids with zero-dimensional nodal points in the band structure close to the Fermi energy. These points realise Dirac and Weyl fermions in 3D. The unusual band topology results in fascinating transport properties:  for instance, negative magnetoresistance is related to the chiral anomaly in WSMs \cite{Jia:2016aa,Yan:2017aa,Lv:2021aa}.

The cone-like band structure of DSMs and WSMs results in optical properties with some similarity to the 2D DSM graphene, such as the possibility of absorbing infrared radiation over a wide frequency range  \cite{Nair:2008aa}, with resulting applications in photodetection \cite{Yang:2019ab,Liu:2020ad}, as well as effects owed to the chiral nature of the Weyl fermions \cite{Shao:2015aa,Baum:2015aa,Chan:2017aa}.  A key property with respect to photodetection is the ultrafast electron dynamics \cite{Weber:2021aa}. Optical studies find similarities to graphene \cite{Weber:2015aa,Lu:2017ab,Ishida:2016aa,Weber:2017aa,Zhu:2017ab}, where the Dirac point, if it occurs above the Fermi $E_F$, can serve as a bottleneck for the relaxation of photoexcited carriers \cite{Butscher:2007aa,Tse:2009aa,Wang:2010ac,Johannsen:2013ab,Johannsen:2015aa}. However, DSMs and WSMs are 3D materials, and optical experiments are not $\mathbf{k}$-resolved, so it is not possible to conclusively link the overall dynamics observed in those experiments to a particular part of the bulk Brillouin zone (BZ). 

Time- and angle-resolved photoemission spectroscopy (TR-ARPES) has the potential to address this: It  can probe unoccupied electronic states and reveal their ultrafast electron dynamics as a function of energy and $\mathbf{k}$ \cite{Sobota:2012aa,Johannsen:2013ab,Boschini:2024aa}. However, performing TR-ARPES on WSMs is highly challenging, due to the materials' 3D nature \cite{Caputo:2018aa,Crepaldi:2017ab,Wan:2017aa,Hein:2020ab}. The conventional approach to TR-ARPES uses a single photon energy as a probe; and, in doing so, essentially assumes a 2D character in the material. In the case of a WSM, this assumption is not valid. Here we use a broad range of photon energies from a high harmonic comb, and probe the 3D BZ in a way similar to what is standard procedure when using synchrotron radiation. We find a pronounced photon-energy dependence of the unoccupied band structure and of electron dynamics, which we can explain in terms of the 3D band dispersion.

\begin{figure}
\includegraphics[width=0.5\textwidth]{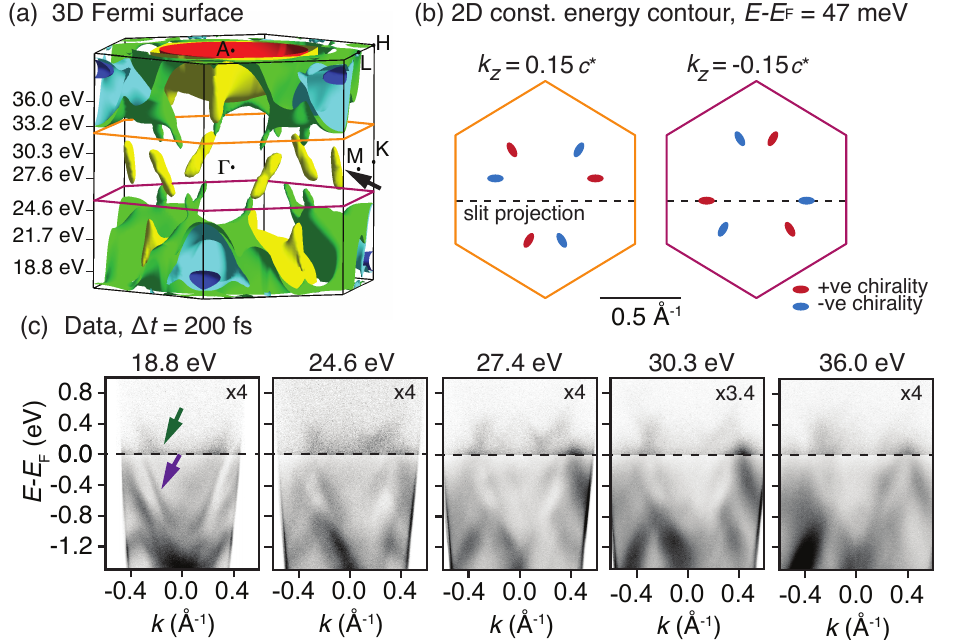}\\
  \caption{(Color online) (a) Bulk Brillouin zone and Fermi surface of PtBi$_2$. The $k_z$ values for the photon energies in our experiment are marked, as well as the planes containing the Weyl points. 
  (b) Two-dimensional cuts through the Brillouin zone at $k_z$ indicated by the corresponding colours in panel (a). The dashed line shows the approximate location of the reported ARPES measurements. The markers represent the location of the Weyl points and the colour their chirality. 
  (c) Dispersion along this dashed line in panel (b) for different probe photon energies following pumping by a 1.46~eV laser pulse at peak excitation. A different grey scale is chosen for the states below and above the Fermi energy, as indicated by insets. Teal (purple) arrows point to the metallic bulk (surface) states.}
  \label{fig:1}
\end{figure}

Trigonal PtBi$_2$ has recently been identified as a type-I WSM \cite{Veyrat:2023aa}. The band structure of the material has been characterised by ARPES \cite{Thirupathaiah:2018aa,Yao:2016aa,Feng:2019aa,Jiang:2020ab,Kuibarov:2024aa}, although the WPs, predicted to be 48~meV above the Fermi level $E_\mathrm{F}$ \cite{Veyrat:2023aa}, are not accessible by this technique. However, it has been possible to detect the topological Fermi arc-type surface states connecting WPs of opposite chirality, providing strong evidence for the WSM character of PtBi$_2$ (for a sketch of the WP positions, see Fig. \ref{fig:1}(a), (b)). A unique feature of the Fermi arc surface states is their superconducting nature below 10~K \cite{Kuibarov:2024aa}. 

PtBi$_2$ single crystals were grown via the self-ﬂux method, as outlined in Refs.~\cite{Shipunov:2020aa,Veyrat:2023aa,Kuibarov:2024aa}. Clean surfaces were prepared \textit{in situ} by cleaving. For TR-ARPES measurements, pump and probe beams were generated from a single 100-kHz laser system \cite{Thire:2023aa}. The 1750-nm output was frequency-doubled and used both for pumping and for driving high-harmonic generation in an argon gas jet. The harmonics were selected by a time-preserving grating monochromator. The pump and probe beam polarisation were perpendicular and parallel to the plane of incidence, respectively. The pulse width was estimated at $\approx$ 75~fs, on the basis of autocorrelation of the pump with itself. The energy resolution was $\approx$ 130~meV, the pump fluence for the time scans was $\approx$ 0.5~mJ\;cm$^{-2}$ and  the sample temperature was $\approx$ 80~K. A FeSuMa spectrometer \cite{Borisenko:2022wd,Majchrzak:2024aa} was used to identify the sample's azimuthal orientation and to locate a single surface termination of type A \cite{Kuibarov:2024aa}. The detailed spectra shown in this paper were then measured with a hemispherical electron analyser. The sample was tilted by $\approx$5$^{\circ}$, in order to place the captured angular range approximately at the positions of the WPs (see sketch in Fig. \ref{fig:1}(b)). Strictly speaking, the required tilt for exactly passing through the WPs depends on the photon energy; but, due to broadening effects, no major changes could be observed when slightly changing the detection direction.  

The band structure of PtBi$_2$ was calculated for the experimental $P31m$ crystal structure \cite{Shipunov:2020aa}  by the density functional theory (DFT)-based linear muffin-tin orbital (LMTO) method as implemented in the PY LMTO code \cite{Antonov:2004aa}. Relativistic effects were accounted for by solving the Dirac equation in atomic spheres. The exchange-correlation potential was approximated according to Ref.  \cite{Perdew:1996aa}.

In our experiment, the bands around the WPs are transiently occupied by the pump pulse and become accessible to TR-ARPES. However, measuring the band dispersion is still challenging because it requires both placing the detector at the correct position in $\mathbf{k}_\parallel = (k_x, k_y)$ and choosing the correct $k_\perp$. In the simplified picture of free electron final states \cite{Plummer:1982aa}, this can be achieved by selecting the photon energy corresponding to the desired $k_\perp$. Tuning the photon energy over a wide range is standard for synchrotron radiation-based ARPES \cite{Hofmann:2002aa} but is challenging when working with lasers. Here we overcome this by using the full frequency comb generated in the gas jet, giving a spectrum of photon energies covering $k_\perp$ throughout the BZ, as  shown in Fig. \ref{fig:1}(a). Note, however, that $k_\perp$ ceases to be a good  quantum number in ARPES, due to the short inelastic mean free path of the photoelectrons. One thus expects some broadening around the desired value of $k_\perp$. 

The $k_\perp$ dependence of the occupied PtBi$_2$ band structure has been explored using variable photon energy ARPES in Ref.~\cite{Veyrat:2023aa,Kuibarov:2024aa}, along with a determination of the approximate photon energies to reach the A and $\Gamma$ high symmetry points in the BZ along normal emission (19 and 29~eV, respectively). Fig.~\ref{fig:1}(c) shows measurements for both the occupied and the normally unoccupied states after pumping the sample with a 1.46~eV laser pulse for selected photon energies. The photoemission intensity is shown as a function of binding energy and $k_\parallel$, approximately along the dashed line in the BZ of Fig.~\ref{fig:1}(b) and at a delay of $\Delta t=200$~fs after the arrival of the pump pulse. We observe a hot electron distribution populating the states up to $\approx$0.5~eV above $E_\mathrm{F}$. The probe photon energy is scanned through the comb of generated harmonics. Only data acquired at selected probe energies are shown in the figure; for other available energies, see Fig. \ref{fig:1}(a), and for a full data set, see the Online Supplementary Material \cite{supp}. The dispersion is modified heavily upon changing the probe photon energy. At a probe energy of 18.8~eV, and with the cut shown here being close to the A-L direction, a metallic bulk band crosses the Fermi level at $k_\parallel \approx$-0.2~\AA$^{-1}$ and can be observed up to $\approx$0.5~eV above $E_\mathrm{F}$ (marked by a teal arrow). When increasing the photon energy, the single band starts to turn into a broad cone-like structure at $k_\parallel \approx$-0.3~\AA$^{-1}$ and a similar one at $k_\parallel \approx$0.2~\AA$^{-1}$. As we shall see below, the location and dispersion of these cone-like structures are consistent with the predicted electronic structure around the WPs. Interestingly, the metallic surface state observed at h$\nu=$18.8~eV (marked by a purple arrow) \cite{Yao:2016aa,Thirupathaiah:2018aa} remains essentially unexcited.

\begin{figure}
\includegraphics[width=0.5\textwidth]{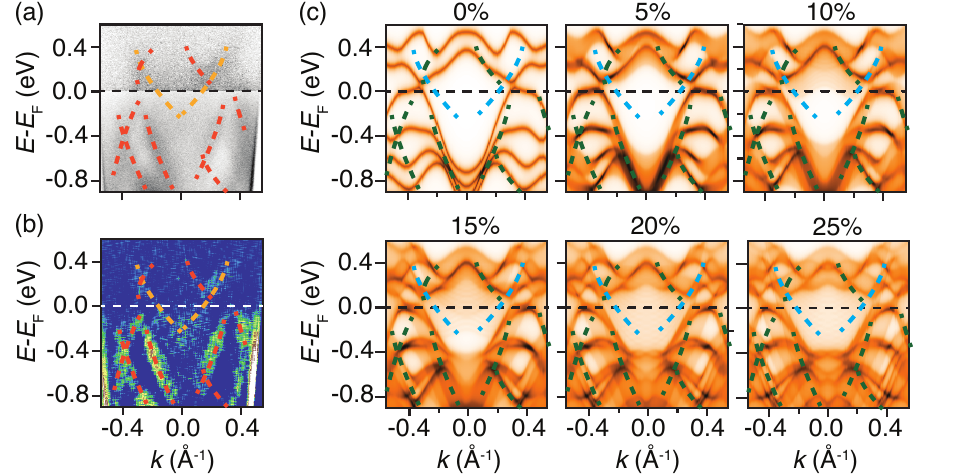}\\
  \caption{(Color online) (a) Spectrum taken at $h\nu = 24.6$~eV, at $\Delta t = 200$~fs. Regions of clear photoemission intensity are emphasised by dashed lines. (b) Second derivative plot of the spectrum in a). 
  (c) DFT calculations with different extents of Gaussian smearing in the bulk Brillouin zone along $k_\perp$, in units of the reciprocal lattice vector $c^{\ast}$. The positions of dashed lines from panel a) have been adjusted to match the calculated features.}
  \label{fig:2}
\end{figure}

A better understanding of the observed features can be achieved by comparing to bulk DFT calculations. In order to account for the uncertainty in  $k_\perp$, we introduce a Gaussian smearing across a range of $k_\perp$ values in the calculations. In Fig.~\ref{fig:2}, the effect of this smearing is systematically investigated starting from data obtained at photon energy corresponding to $k_z \approx 2.81 c*$. Panels (a) and (b) show the spectrum and its second derivative. Some locations with clear photoemission intensity are marked by dashed lines. Fig.~\ref{fig:2}(c) shows the corresponding calculations with the dashed lines from panel (a) adjusted to fit the calculations. The weak electron pocket inside a wide gap (marked by colours different from the other dashed lines) stands out as not being present in the calculations. Its presence can be explained by $k_\perp$ smearing. Starting from a calculation strictly at the $k_\perp$ value corresponding to the measurement, an increasingly wide Gaussian smearing in $k_\perp$ is performed. The full width at half maximum of the $k_\perp$ Gaussian broadening in units of the BZ size $c^{\ast}$ is given for each panel. In order to account for photoemission intensity in the region of the electron pocket, a $k_\perp$ smearing of at least 15\% is required.

\begin{figure}
\includegraphics[width=0.5\textwidth]{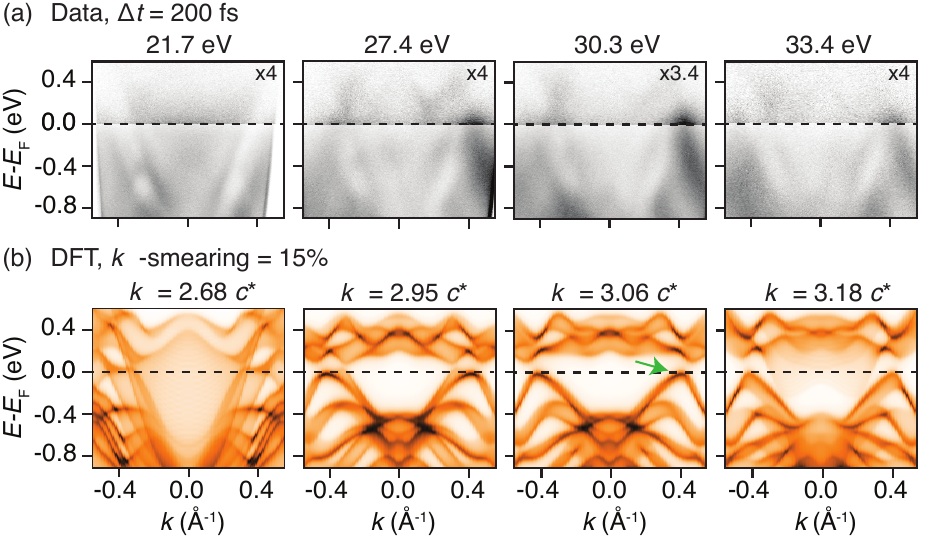}\\
  \caption{(Color online) (a) TR-ARPES spectra collected at different photon energies, $h\nu$. The cut through the momentum space corresponds to the dashed line in Fig.~\ref{fig:1}(b). A different color scale is chosen for the states below and above the Fermi energy, as indicated by insets.
  (b) Density functional theory calculations taken along the same line in $k$-space and with a Gaussian $k_\perp$-broadening corresponding to 15\% of the BZ. The arrow marks the band forming the hole pocket also marked by an arrow in Fig. \ref{fig:1}(a).}
  \label{fig:3}
\end{figure}

Fig.~\ref{fig:3} shows a corresponding comparison of experimental data and Gaussian-broadened calculations for some of the other photon energies explored. As discussed in Ref. \cite{Kuibarov:2024aa}, these calculations are not expected to give a perfect match to the experiment but they do show a good qualitative agreement given the small energy scale. 

\begin{figure}
\includegraphics[width=0.5\textwidth]{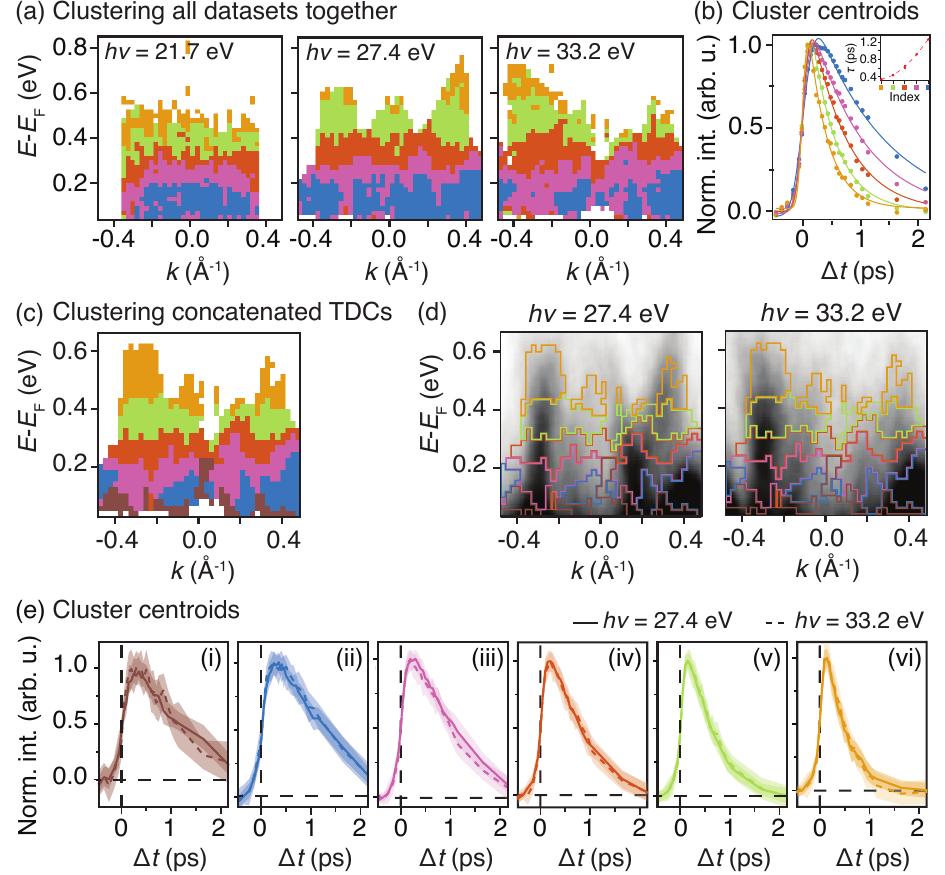}\\
  \caption{(Color online) Results for clustering the data for three photon energies combined using $k$-means ($k=5$, see also Ref. \cite{long_paper}). (a) Cluster index distribution in $(E,k)$ for each photon energy. Areas with the same colour belong to the same cluster and thus show similar dynamics, i.e., time distribution curves (TDCs) with a similar shape. (b) Normalised TDCs averaged over all areas of the corresponding colour. The lines represent a fit with a single exponential decay. The resulting lifetime $\tau$ is plotted in the inset.  (c)  Result for $k$-means clustering of the \emph{concatenated} TDCs in the $h\nu=$27.4 and 33.2~eV data sets. (d) Maps outlining the cluster distribution on the photoemission intensity at  $\Delta t = 200$~fs. (e) Corresponding cluster centroids, plotted such that the TDCs for the two photon energies are shown on top of each other.}
  \label{fig:4}
\end{figure}

We proceed to study the ultrafast electron dynamics throughout the BZ. To this end, we have acquired time-dependent data for three photon energies: 21.7, 27.4 and 33.2~eV. These correspond to, respectively, a cut near the A-L-H plane of the BZ, one close to the WPs [as in Fig. \ref{fig:1}(b), right panel], and one at a corresponding $k_z$ on the other side of the $\Gamma$-M-K plane [and thus missing the WPs, as shown in Fig. \ref{fig:1}(b), left panel]. Combined, this represents a 4D data set, the photoemission intensity $I$ as a function of $E, k$, $\Delta t$ and photon energy $h\nu$. We analyse this data in terms of time distribution curves (TDCs), plotting $I(E,k,h\nu)$ as a function of $\Delta t$ in Fig. \ref{fig:4}(b), in order to extract the ultrafast dynamics in a specific location of ($E, k, h\nu$) space. It is challenging to establish systematic trends. The TDCs differ qualitatively throughout the BZ. In some regions, they are well-described by a fast increase and a single exponential decay, while in others the decay appears delayed and more complex. In order to address this, we make use of the $k$-means, an unsupervised machine learning technique \cite{MacQueen:1967aa,Ball:1967aa}, to classify the TDCs according to their similarity in different ($E, k, h\nu$) locations. Our detailed approach is described in the accompanying paper \cite{long_paper}. Here, we give the key results. 

We evaluate TDCs as intensity integrated over a small $(E, k)$ region of interest (ROI) for each $h\nu$, and exclude ROIs in which the intensity fails to reach a certain threshold. We then normalise each TDC to the have a maximum value of 1, and cluster the resulting set of TDCs \emph{in the entire data set of three photon energies} using $k$-means. We choose $k=5$ clusters. Fig. \ref{fig:4}(a) shows the resulting cluster distribution for the three photon energies. The colour of a ROI corresponds a particular cluster index, meaning that the TDC in this particular ROI is more similar to TDCs of the same ROI colour than to those of the other colours. Fig. \ref{fig:4}(b) shows the average TDCs for each cluster (the so-called cluster centroids). Clearly these show characteristic time scales evolving from ``slow'' to ``fast'' decay as the  energy increases (the TDC colours correspond to the cluster colours in panel (a)). This is a consequence of the non-linear Fermi-Dirac distribution in a hot electron gas \cite{Ulstrup:2014aa}. The centroid TDCs are plotted in the inset, fitted with a single exponential, and the lifetime $\tau$ for the different clusters is given. 

A closer inspection of the 21.7~eV map in Fig. \ref{fig:4}(a) reveals that the ``fast'' decays set in at lower energy than for the other photon energies. This is particularly evident for the two highest energy clusters (orange and green). Consider the fastest-centroid TDC (orange), with a single exponential decay time of  $\tau  = (340\pm20)$~fs. For $h\nu=$21.7~eV, the mean energy where this behaviour is observed is (470$\pm100$)~meV whereas it is (600$\pm80$) and (570$\pm80$)~meV for 
$h\nu=$27.4 and 33.2~eV, respectively. 

The faster dynamics for $h\nu=$21.7~eV can be explained by the overall electronic structure of PtBi$_2$. The data in question is collected near the  A-L-H plane of the bulk Brillouin zone in a close proximity to the main Fermi surface elements (see Fig. \ref{fig:1}(a)). The fast dynamics can be ascribed to the dominance of metallic states and the resulting possibility of fast intra-band Auger decays of excited carriers. This is in sharp contrast to the region near the BZ centre / WPs, where the scans for the other two photon energies were taken: The only Fermi surface elements in the $\Gamma$-M-K plane are the yellow ``cigars'' marked by an arrow in Fig. \ref{fig:1}(a). These arise due to very shallow hole pockets. A band forming such a hole pocket is marked by an arrow in Fig. \ref{fig:3}(b).  The hole pockets' existence depends on small details of the calculation. They are absent in the calculation of Ref. \cite{Kuibarov:2024aa}, for instance. More importantly, even if these Fermi surface elements are present, there is still a gap separating the corresponding bands and the higher-lying states (see Fig. \ref{fig:3}(b)), explaining the slower dynamics. Indeed, the behaviour throughout Fig. \ref{fig:3} is perfectly consistent with our observations: There is a strong metallicity seen for $h\nu=$21.7~eV with several bands crossing $E_\mathrm{F}$ and an additional continuum of states resulting from $k_\perp$ smearing. Near the BZ centre, on the other hand, the electronic structure is semiconductor-like with a pronounced gap. 
 
Note that the absolute time scale for the decay at $h\nu=$21.7~eV---a few hundred fs---is still much slower than in a typical metal or in n-doped graphene \cite{Ulstrup:2015aa,Johannsen:2015aa}. However, the $k_\perp$ smearing is likely to make the decay appear slower than it really is. Similarly to the situation in Fig. \ref{fig:2}, photoemission intensity from other $k_\perp$ values needs to be considered, and longer decay times---like those seen at  $h\nu$=27.4 and 33.2~eV---will always mask a faster decay. 

The clustering maps for $h\nu$=27.4 and 33.2~eV in Fig. \ref{fig:4}(a) appear to show a similar dynamics. However, the differences between these two data sets are particularly interesting because the scan at $h\nu$=27.4~eV passes quite close to the WPs whereas the one at 33.2~eV does not, despite being in a similar region of the BZ (the two scans approximately correspond to the dashed lines in Fig. \ref{fig:1}(b)). In order to specifically emphasise the difference between the data sets, we cluster the set of \emph{concatenated} TDCs from the corresponding ROIs in both data sets using $k$-means \cite{long_paper}. Set up like this, $k$-means sorts objects into the same class if their \emph{combined} shape is the same in both data sets. In order to be sorted into different classes, it is now sufficient that the TDC line shape at one photon energy changes while that at the other photon energy remains the same. 

The result of this clustering is shown in Fig. \ref{fig:4}(c)-(e), as cluster distribution over the ROIs, cluster distribution superimposed on the photoemission intensity at $\Delta t = 200$~fs and cluster centroids, respectively. The TDCs corresponding to the cluster centroids are shown such that the TDCs from the two photon energies are plotted together for each cluster. As expected, for most of the clusters we do not find  any difference between the photon energies. There are indications of small differences in clusters (i), (iii) and (vi), pointing towards a slightly faster decay at $h\nu$=33.2~eV. Due to the small number of TDCs in cluster (i), the signal is quite noisy and even for the other two clusters, the differences are well within the standard deviations of the curves. Nevertheless, the more detailed investigation of the situation in cluster (iii) shows that the effect is genuine and that the $(E,k)$ regions around the WPs show a consistently different TDC shape than the corresponding regions for $h\nu$= 33.2~eV \cite{long_paper}. Such a behaviour could be explained by the region close to the WPs only offering a slow decay path whereas the scan at  $h\nu$= 33.2~eV is already sufficiently far away from the $\Gamma$-M-K plane to show some metallic spectral weight due to $k_\perp$ broadening. This picture is consistent with an inspection of Fig. \ref{fig:3}(b).

In conclusion, we used TR-ARPES to investigate the unoccupied electronic structure of the WSM PtBi$_2$, finding indications of the WPs above $E_\mathrm{F}$. The hallmarks of the WSM electronic structure are particularly evident in the electron dynamics, which exhibit a slower decay in the gapped region of the BZ, where the WPs occur, than in the region hosting the bulk Fermi surface. Due to the high dimensionality of such data, the discovery of subtle phenomena is greatly eased by using $k$-means, as this does not only find regions of similar dynamics but also yields high-quality spectra when taking the average of all the TDCs in these regions. The benefits and limitations of this technique are further explored in the accompanying paper \cite{long_paper}.

\begin{acknowledgments}
This work was supported by the Independent Research Fund Denmark  (Grant No. 1026-00089B). Access to Artemis at the Central Laser Facility was provided by STFC (Experiment Number 23120004). SA acknowledges DFG through AS 523/4-1.
\end{acknowledgments}

%
%\bibliographystyle{apsrev}
%\bibliography{phref,local}

\end{document}